\documentclass[prl,twocolumn,showpacs,superscriptaddress]{revtex4}
\usepackage{amssymb}
\usepackage{graphicx,color}
\begin{document}

\title{Critical Behaviour of the Number of Minima of a Random Landscape at the Glass Transition Point and the Tracy-Widom distribution}
\author{Yan V. Fyodorov}
 \affiliation{Queen Mary University of London, School of Mathematical Sciences, London E1 4NS, United Kingdom}
 \author{Celine Nadal}
 \affiliation{Rudolf Peierls Center for Theoretical Physics,
University of Oxford, OX1 3NP, UK and All Souls College, Oxford, OX1 4AL, UK}

\date{\today}

\begin{abstract}

We exploit a relation between the mean number ${\cal N}_{m}$ of minima of random Gaussian surfaces
and extreme eigenvalues of random matrices to understand the critical behaviour of ${\cal N}_{m}$ in the simplest glass-like
transition occuring in a toy model of a single particle in $N$-dimensional random environment, with $N\gg 1$. Varying the control parameter
$\mu$ through the critical value $\mu_c$ we analyse in detail how ${\cal N}_{m}(\mu)$ drops from being exponentially large
in the glassy phase to ${\cal N}_{m}(\mu)\sim 1$  on the other side of the transition. We also extract a subleading behaviour of ${\cal N}_{m}(\mu)$ in both glassy and simple phases. The width $\delta{\mu}/\mu_c$ of the critical region is  found to scale as $N^{-1/3}$ and inside that region ${\cal N}_{m}(\mu)$ converges to a limiting shape expressed in terms of the Tracy-Widom distribution.

\end{abstract}
\pacs{
05.40.-a,75.10.Nr}

\maketitle

Understanding statistical structure of stationary points (minima, maxima and saddles) of random landscapes and fields of various types is a rich problem of intrinsic current interest in various areas of pure and applied mathematics~\cite{math1,math2}. It
  also keeps attracting steady interest in theoretical physics community over more than fifty years~\cite{sea,halp1,Freund1995,Lennard-Jones,my2004,KleinAgam}, with recent applications to statistical physics \cite{spinglass1,spinglass2}, string theory~\cite{string} and cosmology~\cite{cosm1,cosm2}. For a  landscape described by a sufficiently smooth
random function ${\cal H}({\bf x})$  of $N$ real variables ${\bf
x}=(x_1,...,x_N)$ the problem of counting all stationary points amounts to finding solutions of
the simultaneous stationarity conditions  $\partial_k {\cal H}=0$
for all $k=1,...,N$, with $\partial_k$ standing for the partial
derivative $\frac{\partial}{\partial x_k}$.  Finding the total number
$N^{(D)}_{tot}=\int_D \rho({\bf x}) \, d^N{\bf x}$ of stationary points in any spatial domain $D$
amounts to knowing the corresponding density $\rho({\bf x})$ essentially given by the Jacobian associated with the Hessian $\partial^2_{k_1,k_2}
{\cal H}$ of the random surface at stationary points. In particular, the mean value of such density is given by the so-called Kac-Rice formula
\begin{equation}\label{KR}
\rho_{av}({\bf x})=\left\langle |\det{\left(\partial^2_{k_1,k_2}
{\cal H}\right)}| \prod_{k=1}^N\delta(\partial_k {\cal
H})\right\rangle
\end{equation}
where $\delta(x)$ stands for the Dirac's $\delta-$ function and brackets here and henceforth denote the ensemble average.
For a general random surface the problem of evaluating the averages involving the modulus of the Jacobian is rather difficult and no  efficient technique seems to be known to perform the task~\cite{my2005}.
However, as was first noticed in~\cite{my2004} such calculation can be indeed completed for  Gaussian fields ${\cal H}({\bf x})$ such that their covariance structure $\left\langle
{\cal H}({\bf x}_1) {\cal H}({\bf x}_2)\right\rangle $ depends
only on the Euclidean distance $|{\bf x}_1-{\bf x}_2|$ and is therefore invariant under rotations.
For such fields it turns out to be possible to reduce (\ref{KR}) to evaluating the mean density of eigenvalues of the Gaussian Orthogonal Ensemble (GOE) of real $N\times N$ random
matrices
for which closed-form expression is well-known~\cite{Mehta}. This observation and its further ramifications proved to be very useful
in estimating the probability density of the highest maximum of such surfaces~\cite{math2}, for
 obtaining detailed information about large-$N$ asymptotics of $N^{(D)}_{tot}$ in spherical spin-glass models~\cite{Auffinger}
and in counting stationary points of random superposition of eigenfunctions of Laplacian operator on high-dimensional manifolds~\cite{Nicolaescu}.

Among various possible types of random Gaussian landscapes we will concentrate on arguably the simplest, yet nontrivial model
\begin{equation}\label{fundef}
{\cal H}=\frac{\mu}{2}\sum_{k=1}^N x_k^2+V(x_1,...,x_N)
\end{equation}
where $\mu>0$ is the main control parameter
and $V({\bf x})$ is a random mean-zero Gaussian-distributed field
characterized by a particular (translational invariant) covariance structure:
\begin{equation}
\langle V(\mathbf{x}) V(\mathbf{y})\rangle = N\: f\left(\frac{1}{2 N} (\mathbf{x}-\mathbf{y})^2\right)
\end{equation}
where $f(x)$ is any smooth function which in this paper we consider suitably decaying at infinity.

   Looking at 
   (\ref{fundef}) as a certain random energy surface, one can associate with it the equilibrium statistical mechanical model of a single particle equilibrated by thermal forces as described by the corresponding equilibrium Boltzmann-Gibbs measure $p_{\beta}({\bf x})=e^{-\beta V({\bf x})}/Z(\beta)$, with $\beta=T^{-1}$ being the inverse temperature and $Z(\beta)$ being the associated partition function. The model can be then studied in the framework of the replica trick method,
see \cite{MP,Engel,FS}, though nowadays a rigorous mathematical treatment of  models of such type is possible as well, see e.g. \cite{Klim}.
The analysis reveals that for $\mu<\mu_c=\sqrt{f''(0)}$ the model exibits for $N\to \infty$ a well-defined thermodynamic transition to a glassy
phase with single-step broken replica symmetry. In particular, in the latter phase confined to the lower-temperature region $T<T_c(\mu)$ in the $\mu-T$ plane the thermodynamic expectation value of the particle displacement $\frac{1}{N}\int {\bf x}^2 p_{\beta}({\bf x})\,d{\bf x}$ becomes $T-$ independent, i.e. particle remains "frozen" below the transition line. It is therefore conventional to consider such model as a toy system describing the glass-like freezing transition. The transition temperature $T_c(\mu)$ tends to zero when $\mu\to \mu_c$, so that for $\mu>\mu_c$ even for $T=0$ the system remains in the
 replica symmetric phase. On the other hand for $\mu<\mu_c$ the replica symmetry at zero temperature is broken. In this way the model
provides a rather unique possibility of studying zero-temperature glass transition by varying the control parameter $\mu$.

As the statistical mechanics at $T\to 0$ is expected to be dominated by minima in the energy landscape, the existence of the zero-temperature phase transition suggests that the number of landscape minima $\mathcal{N}_m$ in the two phases should be qualitatively different. Namely, it is expected that $\mathcal{N}_m(\mu)\propto \exp{\left(N\Sigma_m(\mu)\right)}$, with $\Sigma_m(\mu)>0$ for $\mu<\mu_c$
when the random part of $\mathcal{H}$ dominates, but drops to a subexponential value for $\mu>\mu_c$
when the harmonic part of $\mathcal{H}$ dominates and tends eventually to $\mathcal{N}_m=1$ for $\mu\to \infty$.
In such a context, it is natural to call the quantity $\Sigma_m(\mu)>0$ the {\it complexity of minima}.

Even for such a simple model (\ref{fundef}) understanding the statistical properties of the number of minima $\mathcal{N}_m(\mu)$ is a difficult, and to large
extent open problem. Some progress is possible if one concentrates on studying the simplest informative quantity, the mean
value $\left\langle \mathcal{N}_m(\mu)\right\rangle$. And even analysing that mean as a function of $\mu$
is not at all trivial, and is actually the subject of the present Letter. To that end let us recall that the original paper \cite{my2004}
concentrated on analysing even simpler quantity, the mean  of the total number of stationary points $N^{(D)}_{tot}(\mu)$, and on extracting the associated {\it cumulative complexity} $\Sigma_t\left(\mu\right)$. Further developments of the method proposed in~\cite{BrayDean}, and applied to the present model in~\cite{my2007} allowed to evaluate the complexities associated with the mean number of stationary points with any {\it extensive} index $I=\alpha N \gg 1$ with $\alpha >0$, where the index $I$ is defined as the number of negative eigenvalues of the Hessian at the stationary point.  One then can consider the formal limit $\alpha\to 0$  and in this way extract the supposed complexity of minima which turned out to be given by \cite{my2007}
\begin{equation}\label{eq:NmLogeq}
 \Sigma_m\left(\mu\right)=-\ln\left(\frac{\mu }{\mu_c}\right)-
\frac{\mu ^2}{2 \mu_c^2}+\frac{2 \mu }{\mu_c}-\frac{3}{2}\;\; \textrm{when $\mu < \mu_c$}
\end{equation}
and $\Sigma_m(\mu)=0$  for $\mu>\mu_c$. Note however that $\alpha\to 0$ can not distinguish genuine minima with $I=0$ from any other saddle-points
with non-extensive $0<I\ll N$.

The goal of this letter is to perform the 
accurate evaluation of  $\left\langle \mathcal{N}_m\right\rangle$ directly from the first principles,
and in particular to analyse in great detail the so-called critical regime, i.e. the vicinity of the critical point $\mu=\mu_c$ where
the number of minima drops from its exponentially big value in the glassy phase $\mu<\mu_c$ to a subexponential value for $\mu>\mu_c$.
 Our starting point is the general Kac-Rice expression for the mean number of minima
$\langle\mathcal{N}_m \rangle = \int  \rho_m(\mathbf{x}) d^N \mathbf{x}\:$
with $ \rho_m(\mathbf{x})$ now given by (cf.\ref{KR})
\begin{equation}\label{eq:rhom}
\rho_m(\mathbf{x}) = \left\langle
\left|\det\left(\partial_{i,j}^{2}\mathcal{H}\right)\right|\:
\theta\left(\partial_{i,j}^{2}\mathcal{H}\right)
\: \prod_{k=1}^N \delta\left(
\partial_k \mathcal{H}
\right)\right\rangle\,,
\end{equation}
where the matrix Heaviside step-function $\theta(A)=1$ for positive definite matrices and zero otherwise, which ensures that only strict minima
are contributing to the counting. For the model in question (\ref{fundef}) we can now follow the method of \cite{my2004} and show that
\begin{equation}
 \langle \mathcal{N}_m \rangle=\frac{1}{\mu^N}\:
 \int_{-\infty}^{+\infty} dt\:
\:\sqrt{\frac{N}{2\pi}}\;e^{-N \frac{t^2}{2}} K_N(z_t)
 \end{equation}
 where $z_t=\mu+\mu_c t$  and we introduced
\begin{equation}\label{main}
 K_N(z)=\big\langle
\left|\det\left(z   - M_0\right)\right|\:
\theta\left(z  - M_0\right)\big\rangle_{M_0}
\end{equation}
 where the average now goes over the random matrix $M_0$ taken from the so-called Gaussian Orthogonal Ensemble (GOE)~\cite{Mehta}
 with the probability density $ P(M_0)= C_N \; \exp\left\{-\frac{N}{4 \mu_c^2}\, {\rm tr}M_0^2\right\}$.
Using the $O(N)$ invariance of the GOE measure, we  can introduce the eigenvalues $\lambda_i$ of $M_0$
and in the standard way~\cite{Mehta} find that $K_N(z)=z_N^{-1}\:(2 \mu_c^2/N)^{\frac{N(N-1)}{4}+N}\;\; \tilde{\kappa}_N(y)\;\;\textrm{with $z$ and $y$ related by}\;\;z=y \sqrt{\frac{2 \mu_c^2}{N}}$, the normalization factor $z_N$ given by:
 \begin{equation}
 z_N={\mu_c}^{\frac{N(N+1)}{2}}\,\left(\frac{2}{N}\right)^{\frac{N(N+1)}{4}}\left(\frac{2}{\sqrt{\pi}}\right)^N\,
 (2\pi)^{\frac{N}{2}}\,
 \prod_{j=1}^N \Gamma(1+j/2)
 \end{equation}
and where
 \begin{equation}
\tilde{\kappa}_N(y)=\int_{-\infty}^y d\lambda_1 ... \int_{-\infty}^y d\lambda_N\:
\prod_{i<j} |\lambda_i-\lambda_j| \prod_{i=1}^N (y-\lambda_i) \; e^{-\frac{\lambda_i^2}{2}}
 \end{equation}

In fact in the mathematical literature one frequently uses the "standardized" GOE
defined as the ensemble of real symmetric $N\times N$ matrices $M$ with the measure
$\mathcal{P}(M) \propto \; e^{-\frac{1}{2}\, {\rm tr}M^2}$. To that end define the partition function $Z_N(y)$ as
\begin{equation}
Z_N(y)=\int_{-\infty}^{y} d\lambda_1 ... \int_{-\infty}^{y} d\lambda_N\:
\prod_{i<j} |\lambda_i-\lambda_j|  \; e^{-\frac{\lambda_i^2}{2}}
\end{equation}
By definition, the probability that the maximal eigenvalue
of a standardized GOE matrix  $M$ is smaller than some value $y$
is given by:
\begin{equation}\label{eq:cdfmax}
\mathbb{P}_N(\lambda_{\rm max} \leq y )=\frac{Z_N(y)}{Z_N(\infty)}
\end{equation}
Our key observation is that the function $\tilde{\kappa}_{N}(y)$ is related
to the above cumulative distribution (\ref{eq:cdfmax}) as
\begin{equation}
\frac{d Z_N(y)}{d y}=N\: e^{-\frac{y^2}{2}}\; \; \tilde{\kappa}_{N-1}(y)
\end{equation}
Moreover, a simple change of variables shows that
$Z_N(\infty)=z_N\, \left(\sqrt{\frac{N}{2 {\mu_c}^2}}\right)^{\frac{N(N+1)}{2}}$.
In this way we can express the mean number of minima of our random energy surface as $\langle \mathcal{N}_m \rangle = \left(\frac{\mu_c}{\mu}\right)^N\ \frac{2^{\frac{N+3}{2}}\: \Gamma\left(\frac{N+3}{2}\right)}{\sqrt{\pi}\, (N+1)\,
N^{\frac{N}{2}}}\, I_N(\mu/\mu_c)$ where
\begin{eqnarray}\label{eq:NmfctnTW}
I_N(\mu/\mu_c)=\int_{-\infty}^{+\infty}  e^{\frac{y^2}{2}-\frac{N}{2}\left(y\sqrt{\frac{2}{N}}-\frac{\mu}{\mu_c}\right)^2}\;\\
\times \frac{d}{dy} \left[\mathbb{P}_{N+1}(\lambda_{\rm max} \leq y )\right]\, dy \;\nonumber
\end{eqnarray}
with $\mathbb{P}_{N}(\lambda_{\rm max} \leq y )$ defined in (\ref{eq:cdfmax}). It is appropriate to mention that a closely related formula
appeared also in \cite{Auffinger} in the context of studying the extrema of the random energy surface for the so-called
spherical model of spin glasses.

Main utility of the above observation is in the fact that the behaviour of  the cumulative distribution of the maximal eigenvalue
of GOE matrices for large $N$  was thoroughly studied in various regimes in recent years, starting from the famous work by Tracy and Widom \cite{TW}. In particular, most detailed large-deviation results were obtained in \cite{BorotNadal11} for the left tail and  \cite{BorotNadal12} for the right tail for the regime $N\gg 1$ and $y=s \sqrt{N}$ with fixed $s$. For example, the left tail is given by
 \begin{equation}\label{eq:LeftLargeDevEq}
\frac{d}{dy}\left[\mathbb{P}_N(\lambda_{\rm max} \leq y )\right]\sim
e^{-\Psi_N\left(\frac{y}{\sqrt{N}}\right)}\;\; \textrm{for}\;\; y<\sqrt{2 N}\;
\end{equation}
where $y-\sqrt{2N}= O(\sqrt{N})$ and \cite{DeanMajumdar},\cite{BorotNadal11}
\begin{eqnarray} \nonumber
\Psi_N(s)&=&N^2 \psi_-(s)-N \Phi_1(s)+\phi_1 \ln N+\Phi_2(s)\\ \nonumber
\psi_-(s)&=&\frac{s^2}{3}-\frac{s^4}{108}-\left(\frac{s^3}{108}+ \frac{5 s}{36}\right)\sqrt{s^2+6}\\  &-&
\frac{1}{2}\ln\left[\frac{s+\sqrt{s^2+6}}{3 \sqrt{2}}\right]
\end{eqnarray}
and explicit expressions for  $\Phi_{1,2}(s)$ and $\phi_1$ are rather long and can be found in \cite{BorotNadal11}.
On the other hand, the right large deviation tail is given by \cite{BorotNadal12}
\begin{equation}\label{eq:RightLargeDevEq}
\frac{d}{dy}\left[\mathbb{P}_N(\lambda_{\rm max} \leq y )\right] \sim
\frac{e^{-N \psi_+(s)} }{2 \sqrt{\pi } \left(-2+s^2\right)^{1/4} \sqrt{s+\sqrt{-2+s^2}}} \nonumber\\
\end{equation}
for $s=\frac{y}{\sqrt{N}}\;\;\textrm{and}\;\;y>\sqrt{2 N}$ where \cite{MajumdarVergassola09},\cite{BorotNadal12}
\begin{equation}
\psi_+(s)=\frac{s^2}{2}\sqrt{1-\frac{2}{s^2}}+\ln\left[\sqrt{\frac{s^2}{2}}-\sqrt{\frac{s^2}{2}-1}\right]
\end{equation}
These results allow us to get the behaviour of the mean number of minima by extracting the appropriate asymptotic from the integral (\ref{eq:NmfctnTW}) for large $N$ via the saddle point method.
Using (\ref{eq:RightLargeDevEq}), we find for $\mu>\mu_c$ the following equivalent:
\begin{equation}\label{eq:NmeqRight}
\langle \mathcal{N}_m \rangle \sim 1\;\; \textrm{when}\;\; \mu > \mu_c
\end{equation}
which is much more precise than just the statement of vanishing complexity.
It turns out we just have on average
 {\it one single minimum}
 not only for $\mu\gg \mu_c$ as one might naively expect, but immediately for all $\mu > \mu_c$.
 Similarly, (\ref{eq:LeftLargeDevEq}) implies  that $\langle \mathcal{N}_m \rangle \sim e^{S_N(\mu/\mu_c)}$ for $\mu<\mu_c$
 where
\begin{eqnarray}\label{eq:NmeqLeft}
 S_N\left(\frac{\mu}{\mu_c}\right)&=&N \;\Sigma\left(\frac{\mu}{\mu_c}\right)+\sqrt{N}\Sigma_1\left(\frac{\mu}{\mu_c}\right) \\
&+&N^{1/4}\; \Sigma_2\left(\frac{\mu}{\mu_c}\right)\,+\Sigma_3\left(\frac{\mu}{\mu_c}\right)\,,
\end{eqnarray}
with the leading term $\Sigma\left(y\right)=-\ln\left(y\right)-\frac{y^2}{2}+2 y-\frac{3}{2}$ coinciding with the complexity of minima Eq. (\ref{eq:NmLogeq}) found in \cite{my2007} and subleading terms given by $\Sigma_1\left(y\right) =\frac{4}{3}  \sqrt{2}  \left(1-y\right)^{3/2}$, $\Sigma_2\left(y\right) =-\frac{2}{3} 2^{3/4}  \left(1-y\right)^{3/4}$ and
\begin{eqnarray}
\Sigma_3(y) =-2+2y+\frac{1}{4}y^2+\frac{137 \ln 2}{96}+\frac{\ln\pi }{2}+\\
+\frac{23}{32} \ln\left([1-y]N^{1/3}\right)+\frac{1}{2} \zeta'(-1)
\end{eqnarray}
It is easy to check that $\Sigma\left(\mu\right)\propto (\mu_c-\mu)^3$ when approaching the transition point as was already noticed in \cite{my2007}. This should be contrasted with the cumulative complexity for all stationary points (instead of minima)
vanishing as  $\Sigma_t\left(\mu\right) \propto (\mu_c-\mu)^2$~\cite{my2004}.
The difference is significant as it has implications for the width of the so-called "transition region" $|\mu-\mu_c|$ where the two phases of the system become indistinguishable for large but finite $N\gg 1$. Here we argue that it is the complexity of minima which provides the correct scaling.
Indeed, as shown in \cite{FS} the difference $\Delta F$ between the zero-temperature free energies $F=-\lim_{T\to 0} T\ln{Z(\beta)}$ of the replica-symmetric solution and one with broken replica symmetry is of the order of $N(\mu_c-\mu)^3$ close to the transition point.
Thus thermodynamically the two phases cease to be distinguishable precisely in the same region when the leading term $N \;\Sigma\left(\frac{\mu}{\mu_c}\right)$ in the $\ln{\mathcal{N}_m(\mu)}$
 becomes of the order of unity. One may consider this as an independent
 confirmation of the thermodynamic relevance of minima, rather than of the totality of stationary points for the glass transition.

Moreover, we further see that for $|\mu-\mu_c|\sim N^{-1/3}$ not only the leading term $N \;\Sigma\left(\frac{\mu}{\mu_c}\right)$ but
all the subleading terms  in (\ref{eq:NmeqLeft}) become simultaneously of order of unity. This fact implies that in the transition region where $\left(\frac{\mu}{\mu_c}-1\right)N^{1/3}=\delta$
 is of the order of unity the mean number of minima tends to a $N-$ independent function of the
scaling variable $\delta$. To find that limiting shape we use the celebrated Tracy-Widom law \cite{TW} for the probability of the maximal eigenvalue
of the standardized GOE matrix:
\begin{equation}\label{TracyWidom}
\mathbb{P}_N\left(\frac{\lambda_{\rm max}-\sqrt{2 N}}{N^{-\frac{1}{6}}/\sqrt{2}}\leq x\right)\sim \mathcal{F}_1(x)\;\;\;
\textrm{as $N\to\infty$}
\end{equation}
where $\mathcal{F}_1(x)$ is a special solution of the Painleve II equation. Using this fact  we can rewrite  (\ref{eq:NmfctnTW}) as
\begin{eqnarray}
I_N(\mu/\mu_c)&=&\int e^{h_N(x)}  dx \;,\\
h_N(x)&=&\frac{y_x^2}{2}-\frac{N}{2}\left(y_x\sqrt{\frac{2}{N}}-\frac{\mu}{\mu_c}\right)^2
+\ln\mathcal{F}_1{\,'}(x)\;\;\;\;\;\;
\end{eqnarray}
and $y_x=\sqrt{2(N+1)}+x \frac{(N+1)^{-\frac{1}{6}}}{\sqrt{2}}$.
For large $N$, the integral computed by the saddle point method is dominated by the vicinity of
$x^*$ such that
\begin{equation}\label{xstar}
-\frac{d}{dx}\left[\ln \mathcal{F}_1^{\,'}(x^*)\right]=\delta
\end{equation}
We thus get
$\langle \mathcal{N}_m\rangle \sim \mathcal{N} (\delta)$
in the transition region $\frac{\mu}{\mu_c}=1+\delta N^{-\frac{1}{3}}$
where the  limiting shape $\mathcal{N} (\delta)$ is given by
\begin{eqnarray}\label{eq:NmeqTW}
\mathcal{N} (\delta) =
\sqrt{\frac{2\pi}{-\frac{d^2}{dx^2}\left[
\ln \mathcal{F}_1^{\,'}(x)
\right]\Big|_{x^*}}}\; 2\; \mathcal{F}_1^{\,'}(x^*)\; \;e^{x^* \delta-\frac{\delta^3}{3}}\;\;
\end{eqnarray}
whee $x^*$ is the solution of (\ref{xstar}). It is not difficult to check that the scaling function
(\ref{eq:NmeqTW}) which is one of the main results of this work matches smoothly the two regimes
$\mu<\mu_c$ cf (\ref{eq:NmeqLeft}) and $\mu>\mu_c$ cf (\ref{eq:NmeqRight})
as shown in the figure:

\begin{figure}[h!]
\includegraphics[width=0.45\textwidth]{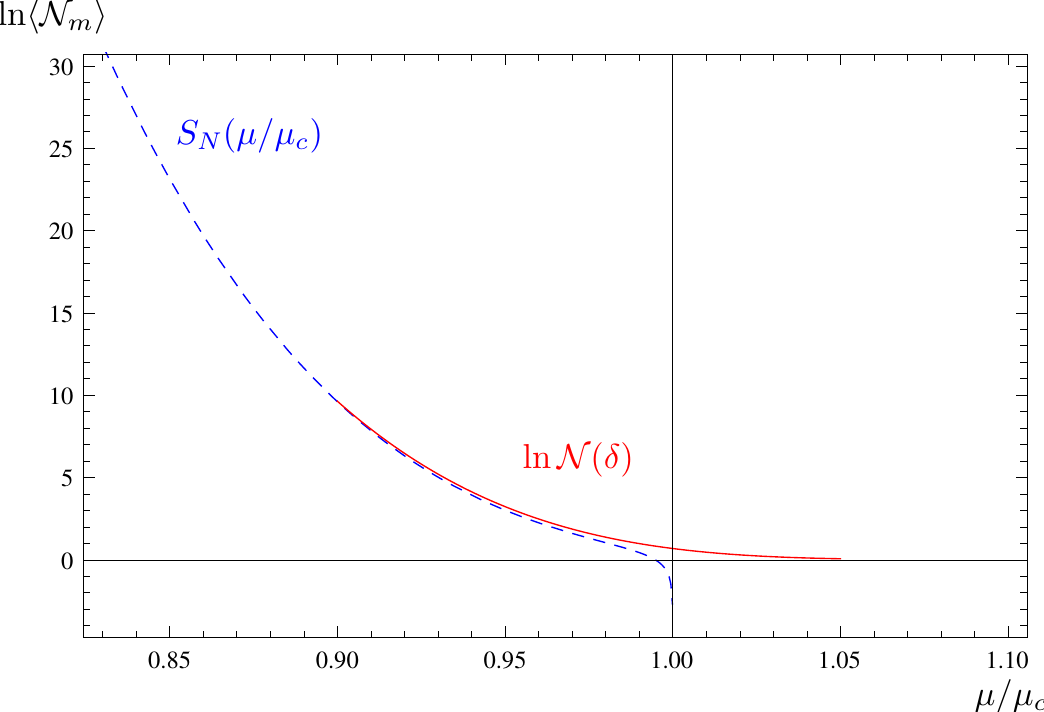}
\caption{Plot of the shape of the (log of the) mean number of minima $\langle \mathcal{N}_m\rangle$
for $N=10000$
(a) for $\mu < \mu_c$:  $\langle \mathcal{N}_m\rangle \sim e^{S_N}$ (blue dashed line)
(b) in the transition region $\frac{\mu}{\mu_c}=1+\delta N^{-1/3}$ (red solid line)
(c) for $\mu >\mu_c$: $\langle \mathcal{N}_m\rangle \sim 1$ (black solid line)}
\label{fig}
\end{figure}

%

In conclusion, we investigated in a great detail the mean number of minima of the random energy function in a toy model of the glass transition, and determined the precise scaling form of that number at the transition region between the two phases. It turned out to be related to the Tracy-Widom distribution well-known in the random matrix theory.  Note that such distribution was also shown to describe fluctuations of transition temperature in mean-field spin glasses\cite{TWSK}.
 We hope that similar methods could prove useful in the problem of understanding the ground states fluctuations of glassy systems, which attracts growing interest in recent years \cite{extremesinglasses}.


\end{document}